\documentclass[journal]{IEEEtran}
\usepackage{amssymb,amsmath,epsfig,psfrag,epstopdf,enumerate}

\usepackage[english]{babel}
\usepackage{color}
\usepackage{graphicx}

\usepackage[normalem]{ulem}
\usepackage{varioref}
\usepackage{balance}
\usepackage{subfig}

\newtheorem{definition}{Definition}

\usepackage{fancyhdr}
\fancypagestyle{firstpage}{
\fancyfoot[C]{This work has been submitted to the IEEE for possible publication.  \\Copyright may be transferred without notice, after which this version may no longer be accessible.}
}

\begin{document}


\title{The Golden Quantizer:\\The Complex Gaussian Random Variable Case}

\author{
\IEEEauthorblockN{Peter Larsson \emph{Student
Member, IEEE},  Lars K. Rasmussen \emph{Senior
Member, IEEE},\\ Mikael Skoglund, \emph{Senior
Member, IEEE}}%
\thanks{The authors are with the ACCESS Linnaeus Center and the School of
Electrical Engineering at  KTH Royal Institute of  Technology, SE-100 44
Stockholm, Sweden. (pla@kth.se)}
}
\maketitle
\thispagestyle{firstpage}

\begin{abstract}
The problem of quantizing a circularly-symmetric complex Gaussian random variable is considered. For this purpose, we design two non-uniform quantizers, a high-rate-, and a Lloyd-Max-, quantizer that are both based on the (golden angle) spiral-phyllotaxis packing principle. We find that the proposed schemes have lower mean-square error distortion compared to (non)-uniform polar/rectangular-quantizers, and near-identical to the best performing trained vector quantizers. The proposed quantizer scheme offers a structured design, a simple natural index ordering, and allow for any number of centroids.
\end{abstract}


\section{Introduction}
\IEEEPARstart{Q}{uantization} of a complex (bivariate, or 2D) Gaussian random variable (r.v.) has been studied in several works.  Pearlman, in \cite{Pearlman79}, Bucklew and Gallagher, in \cite{BucklewGall79}, as well as Wilson, in \cite{Wilson80}, studied the mean-square error (MSE) distortion measure for non-uniform (and uniform) polar-, and rectangular-, quantizers for the complex Gaussian r.v. case. In \cite{Knagenhjelm93}, Knagenhjelm studied vector quantizers (VQs) trained on iid 2D-Gaussian data, and produced, (as far as the authors know), the best performing 2D-Gaussian quantizers to date.

Recently, in \cite{Larsson17a, Larsson17b}, the idea of golden angle modulation (GAM) was proposed. The basic idea for signal constellation design was inspired by spiral phyllotaxis (SP) packing, e.g. observed for the scales on a pine cone.  In this letter, we propose GAM-related non-uniform \textit{golden (angle) quantization} (GQ) designs for quantization of a complex Gaussian r.v. We propose, and develop, a general, a (fixed-, and entropy-coded-) high-rate, and a constrained Lloyd-Max GQ design. We numerically study their MSE-distortions vs. rate, and compare with rectangular- and polar-designs, Knagenhjelm's VQ designs, as well as with the Shannon rate-distortion function.

\section{Golden (Angle) Quantizer}
\label{sec:GQ}
The general golden angle quantizer design (or golden quantizer in short, used henceforth) builds on the use of the golden angle (or golden ratio) for phase rotations of consecutive centroids, but also adapting the centroid magnitudes to attain low distortion. The design is given below.
\begin{definition} (Golden quantizer)
\label{def:Def5p5d1}
The complex valued centroid amplitudes are
\begin{align}
x_n&=r_n\mathrm{e}^{i 2\pi \varphi n}, \, n\in\{1,2,\ldots,N\},
\end{align}
where $r_n$ is the radius of centroid $n$, $2\pi\varphi$ denotes the golden angle in radians, and $\varphi=(3-\sqrt{5})/{2}$.
\end{definition}

We let $r_{n+1}>r_n$ for an increasing spiral winding. A centroid, is located (the irrational number) $\varphi\approx0.382$ turns (or $137.5$ degrees) relative to the previous centroid. Replacing $\varphi$ with $k\pm(3-\sqrt{5})/2$, $k\in \mathbb{Z}$, (e.g. $(1+\sqrt{5})/2$, the golden ratio), gives an equivalent design. 

The two key insights are: i) observing that SP packing offers radial shape-flexibility with retained locally-uniform density, and ii) adopting this shape-flexibility to non-uniform quantization of r.v.s with circular-symmetric densities.

\begin{figure}[tp!]
 \centering
 \vspace{-.4 cm}
 \includegraphics[width=9cm]{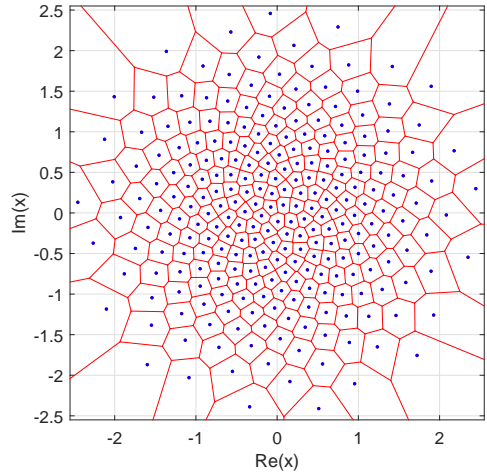}
 \caption{Voronoi-diagram and centroids for Lloyd-Max-GQ, $N=2^8$, optimized for a complex Gaussian pdf with $\sigma^2=1$.}
 \label{fig:LloydMax_01}
 \vspace{-.35cm}
\end{figure}

\subsection{High-rate Golden Quantizer}
\label{sec:HR}
For the first non-uniform quantizer design, we build on high-rate quantization theory, see e.g. \cite{GrayNeuh98}.
The steps are as follows: The radius $r_n$ is determined by first finding the asymptotically optimal quantization point density. Then, this density is integrated and equated to the number of centroids, $n$, within a circle radius of $r_n$. Finally, $r_n$ is determined. 
So first, the asymptotic optimal quantization point density, with a MSE-distortion measure, is
\begin{align} \lambda(x^k)=N\frac{f(x^k)^{\frac{k}{k+r}}}{\int_{\mathbb{R}^k} f(x^k)^{\frac{k}{k+r}} \mathrm{d}x^k},
\end{align}
where $f(x^k)$ is the pdf, $k$ the dimension, and $r$ the distortion norm. For the complex Gaussian r.v., $f(x,y)=\frac{1}{\pi \sigma^2}\mathrm{e}^{-\frac{x^2+y^2}{\sigma^2}}$, and the MSE-distortion, $k=2$ and $r=2$. This gives
\begin{align}
\lambda(x,y)&=\frac{N}{2 \pi \sigma^2}\mathrm{e}^{-\frac{x^2+y^2}{2\sigma^2}}.
\end{align}
From the definition of the point density function, and assuming a circular-symmetric
pdf with radius $r_n$ to point $n$, the condition $\int_{\left\Vert x^2 \right\Vert\ \leqslant r_n} \lambda(x^k) \, \mathrm{d}x^k=n$ must be fulfilled. Hence
\begin{align}
&\iint_{\sqrt{x^2+y^2}\leqslant r_n} \lambda(x,y) \, \mathrm{d}x \, \mathrm{d}y =n\notag\\
&\Rightarrow \int_0^{r_n} \frac{N}{2 \pi \sigma^2}\mathrm{e}^{-\frac{r^2}{2\sigma^2}} 2 \pi r \, \mathrm{d}r =n\notag\\
&\Rightarrow  N(1-\mathrm{e}^{-\frac{r_n^2}{2\sigma^2}})=n \notag\\
&\Rightarrow  r_n= \sigma \sqrt{2\ln \left(\frac{N}{N-n}\right)}.
\label{eq:rn}
\end{align}

\begin{figure}[tp!]
 \centering
 \vspace{-.4 cm}
 \includegraphics[width=9cm]{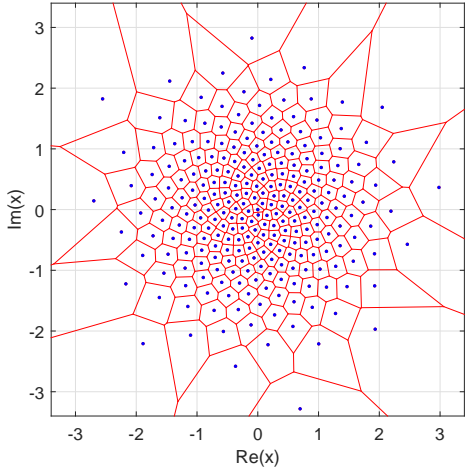}
 \caption{High-rate GQ, $N=2^{8}$.}
 \label{fig:HR}
\vspace{-.35cm}
\end{figure}

The high-rate quantizer design is exemplified in Fig.~\ref{fig:HR}.

\subsubsection{Distortion measure}
An analytical distortion expression for the high-rate design is also of interest. This allows for comparison against the numerically computed distortion for the high-rate case, the Lloyd-Max case, and for the rate-distortion function. The high-rate distortion is defined as
\begin{align}
D \approx C(r,k,G_{k})\frac{1}{N^{\frac{r}{k}}} \left( \int_{\mathbb{R}^2}f_{X^k}(x^k)^{\frac{k}{k+r}}\, \mathrm{d}x^k\right)^{\frac{r+k}{k}},
\end{align}
where $C(r, k, G_k)$ is the normalized moment of inertia, or coefficient of quantization,
\begin{align}
 C(r,k,G_{k})=\frac{1}{k}\frac{1}{V^{\frac{r+k}{k}}}\int_{\mathbb{R}_n}|x^k-\mathcal{Q}(x^{k})|^2 \, \mathrm{d} x^k.
\label{eq:NormMoment}
\end{align}
Eq. \eqref{eq:NormMoment} relies on fixed-sized quantization cell-shapes. However, numerical evaluation of the high-rate design (not shown here) reveals that the numerical normalized moment of inertia $ C(2, 2, G_2)\approx {1}/{12}$, except at the highest quantization indices, where it fluctuates and takes on larger values. This value, ${1}/{12}$, is the same as for the uniform square quantizer.

For the complex Gaussian case, and with an MSE-distortion, we have $k=2$ and $r=2$. Then, the distortion for two dimensions is
\begin{align}
D_\textrm{hr}(k=2)&\approx 2\frac{C(2,2,G_{2})}{N} \left( \int_{\mathbb{R}^2}\sqrt{\frac{\mathrm{e}^{-\frac{x^2+y^2}{\sigma^2}}}{ \pi \sigma^2}}\, \mathrm{d}x \, \mathrm{d}y
\right)^2\notag\\
&=\frac{2\pi\sigma^2}{3N}.
\label{eq:Dhr}
\end{align}
Inserting the rate $R_\textrm{hr}=\log_2(N)$ in \eqref{eq:Dhr}, the high-rate rate-distortion expression becomes
\begin{align}
R_\textrm{hr}=\log_2\left(\frac{2\pi\sigma^2}{3D}\right).
\label{eq:Rhr}
\end{align}
The expression \eqref{eq:Rhr} can be compared with results from rate-distortion theory which state that $R_\textrm{rd}=\log_2(\sigma^2/D)$ for circular-symmetric complex Gaussian r.v. with per dimension variance $\sigma^2/2$, and distortions $D/2$. Hence, the rate difference is $R_\textrm{hr}-R_\textrm{rd}=\log_2\left({2\pi}/{3}\right)\approx 1.067$ [bits].

\subsubsection{Entropy-coded high-rate GQ}
\label{sec:EC}
To illustrate an application of entropy-coded quantizations with GQ, we exemplify with the high-rate GQ design. The (approximate) high-rate centroid probability is $p_n=cf(x_n,y_n)/\lambda(x_n,y_n)$. With $\lambda(x_n,y_n)$ and $f(x_n,y_n)$ expressed in polar coordinates, and inserting $r_n$ from \eqref{eq:rn}, we get $p_n^ \textrm{(echr)}\approx c2(N-n)/N^2$. Normalizing the sum-probability to unity, yields $p_n^\textrm{(echr)}\approx{2(N-n)}/{N(N+1)}$. Using the definition of the entropy, $H=-\sum_{n=0}^{N-1}p_n \log_2(p_n)$, and using the approximation $\sum_{n'=1}^{N}(n'/N) \lg_2(n'/N)\approx \int_0^1 x \log_2(x) \, \mathrm{d}x=1/4\ln(2)=\log_2(\sqrt{\mathrm{e}})/2$ for large $N$, it is straightforward to show that
\begin{align}
H_\textrm{echr}\simeq \log_2(N)-1+\log_2(\sqrt{\mathrm{e}}).
\label{eq:EC}
\end{align}
Because the high-rate design assigns (approximately) the same MSE per cell, the high-rate MSE-distortion is simply $D_\textrm{echr}=N\sum_{n=0}^{N-1}p_n (D_\textrm{hr}/N)=D_\textrm{hr}$, i.e. identical to \eqref{eq:Dhr}.
Combining \eqref{eq:EC} with \eqref{eq:Dhr} gives the high-rate GQ rate-distortion 
\begin{align}
R_\textrm{echr}&=\log_2\left(\frac{\pi \sqrt{\mathrm{e}} \sigma^2 }{3D}\right).
\label{eq:Rechr}
\end{align}
The rate difference between the fixed-, and the entropy-coded-, designs is then $R_\textrm{hr}-R_\textrm{echr}=\log_2(2/\sqrt{\mathrm{e}})\approx 0,279$ [bits].

\subsection{Lloyd-Max Golden Quantizer}
\label{sec:LM}
The high-rate quantizer design is optimal when $N\rightarrow \infty$. However, for limited $N$, the iterative Lloyd-Max quantizer design approach, see e.g. \cite{GrayNeuh98}, can reduce the MSE-distortion further. We minimize the MSE-distortion wrt $r_n$, while constraining the angular distribution of centroids to $2 \pi \varphi n$. The MSE-distortion is expressed as
\begin{align}
D=\frac{1}{2\pi}\sum_{\forall n}\iint_{\mathbb{C}_n}|r\mathrm{e}^{i \phi}-r_n\mathrm{e}^{i \phi_n}|^2f(r) \, \mathrm{d}r \, \mathrm{d} \phi,
\label{eq:DML}
\end{align}
where $\phi_n\triangleq 2\pi \varphi n$, and $\mathbb{C}_n$ is the region of integration (the cells), for centroid $r_n\mathrm{e}^{i \phi_n}$. Let the centroids assume the values $r_n^{(k)}$, and corresponding region for integration $\mathbb{C}_n^{(k)}$, for iteration $k$. We then take the partial derivative of \eqref{eq:DML} wrt $r_n^{(k)}$, equate to zero for an optima, and solve for the updated centroid values $r_n^{(k+1)}$. We get
\begin{align}
r_n^{(k+1)}
&=\frac{\iint_{\, \mathbb{C}_n^{(k)}}r\cos{(\phi-\phi_n)}f(r) \, \mathrm{d}r \, \mathrm{d}\phi}{\iint_{\, \mathbb{C}_n^{(k)}} f(r) \, \mathrm{d}r \, \mathrm{d}\phi}.
\end{align}
To simplify numerical integration, we express the integrals in Cartesian coordinates as
\begin{align}
r_n^{(k+1)}
&=\frac{\iint_{\, \mathbb{C}_n^{(k)}}g(\phi_n,x,y)f\left(\sqrt{x^2+y^2}\right) \, \mathrm{d}x \, \mathrm{d}y}{\iint_{\, \mathbb{C}_n^{(k)}}\frac{1}{\sqrt{x^2+y^2}}f\left(\sqrt{x^2+y^2}\right) \, \mathrm{d}x \, \mathrm{d}y},
\end{align}
where $g(\phi_n,x,y)
\triangleq \frac{x}{\sqrt{x^2+y^2}}\cos{\phi_n} +\frac{y}{\sqrt{x^2+y^2}}\sin{\phi_n}$, and the identity $\cos{(\phi-\phi_n)}=(\cos{\phi}\cos{\phi_n} +\sin{\phi}\sin{\phi_n})$ was used.
Thus, for the golden quantizer approach, with complex Gaussian noise and radial pdf $f(r)=2r\mathrm{e}^{-r^2/\sigma^2}/\sigma^2$, the Lloyd-Max GQ becomes
\begin{align}
r_n^{(k+1)}
&=\frac{\iint_{\, \mathbb{C}_n^{(k)}}(x\cos{\phi_n}+y\sin{\phi_n}) \, \mathrm{e}^{-\frac{x^2+y^2}{\sigma^2}} \, \mathrm{d}x \, \mathrm{d}y}{\iint_{\, \mathbb{C}_n^{(k)}} \mathrm{e}^{-\frac{x^2+y^2}{\sigma^2}} \, \mathrm{d}x \, \mathrm{d}y},\\
\phi_n&=2\pi \varphi n.
\end{align}
In general, for circular-symmetric pdfs, we (may) also impose, $r_{n+1}\geq r_n \forall n$, the condition for a growing spiral.

In Fig. \ref{fig:LloydMax_01}, a Voronoi-diagram, with centroids, for Lloyd-Max optimized GQ is exemplified with $N=2^8$. The quantizer is optimized for a complex Gaussian pdf with variance $\sigma^2=1$. 

\section{Results, Discussion, and Spiral Phyllotaxis}
\label{sec:Results}

\subsection{Numerical Results}
\begin{figure}[tp!]
 \centering
 \vspace{-.4 cm}
 \includegraphics[width=9cm]{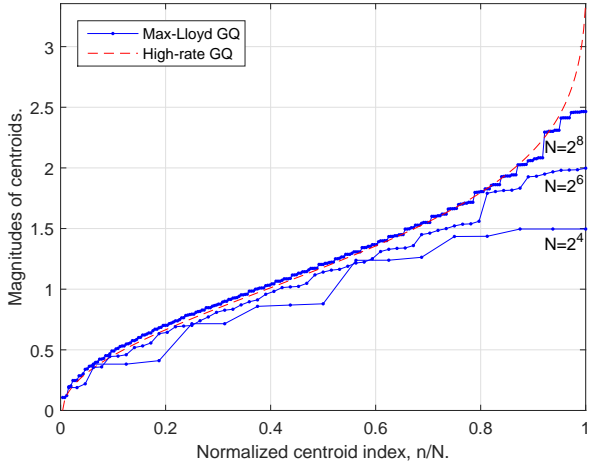}
 \caption{Magnitude vs. normalized centroid index for Lloyd-Max-GQ, and high-rate-GQ, with $N=\{2^4,2^6,2^8\}$, and a complex Gaussian pdf with $\sigma^2=1$.}
 \label{fig:LloydMaxMagn_01}
\vspace{-.35cm}
\end{figure}

The centroid magnitudes vs. the normalized centroid index are plotted for the high-rate-, and the Lloyd-Max-, GQs in Fig.~\ref{fig:LloydMaxMagn_01}. The Lloyd-Max optimized magnitudes approaches the high-rate quantizer magnitudes with increasing $N$. The highest magnitudes for the Lloyd-Max design are limited, and less pronounced, compared with the high-rate case.

\begin{figure}[tp!]
 \centering
 \vspace{-.4 cm}
 \includegraphics[width=9cm]{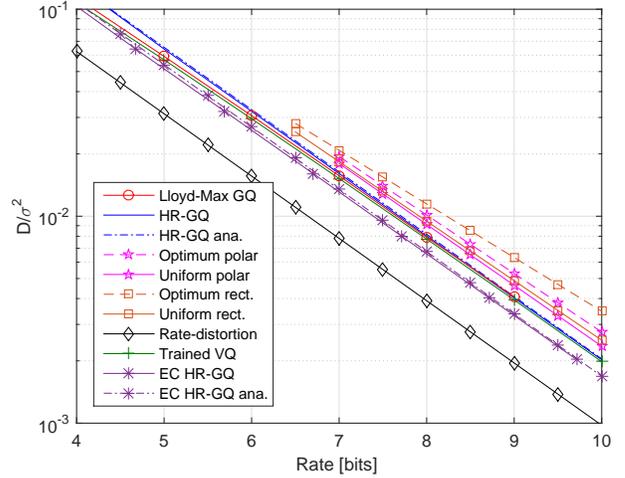}
 \caption{Normalized MSE-distortion vs. rate for Lloyd-Max-GQ, high-rate-GQ, (non)-uniform polar/rectangular-quantizers, trained-VQs, and the rate-distortion function.}
 \label{fig:Perf}
\vspace{-.35cm}
\end{figure}

In Fig.~\ref{fig:Perf}, we plot the performance of the Lloyd-Max-, and high-rate-, golden quantizers. The analytical high-rate GQ distortion expression, \eqref{eq:Dhr}, is also plotted. For comparison, plots for the rate-distortion function, Kragenhjelm's trained VQs \cite[p.15]{Knagenhjelm93}, as well as empirically fitted curves presented in \cite[p.897]{Pearlman79}, for optimal polar, optimal rectangular, uniform polar, and uniform rectangular quantizers, are included.
We find that the Lloyd-Max-optimized- and high-rate-GQ performs better than the polar-, and rectangular-, quantizer cases for relevant rates. However, the trained (best known) VQ designs are (just) slightly better than the Lloyd-Max GQ. Numerical results for the high-rate scheme agrees extremely well with the analytical expression \eqref{eq:Dhr}. As expected, the Lloyd-Max quantizer has lower distortion than high-rate quantizer and the analytical expression \eqref{eq:Dhr}, particularly visible for the lower rates. For increasing rate, the Lloyd-Max quantizer distortion, as well as the trained-VQ, approaches the high-rate quantizer distortion. We also illustrate \eqref{eq:Rechr}, and numerically computed performance, for the entropy-constrained high-rate quantization case. As expected, the entropy-coded design performs better than the fixed design(s). The numerical computed performance also agrees well with \eqref{eq:Rechr}.

\subsection{Discussion}
The most interesting feature of GQ is not necessarily the improved MSE performance. First, the rate $R=\log_2(N)$ is fully-flexible, since $N$ is not restricted to certain integer values, as for polar/rectangular-quantizers. E.g., odd integer rates, $R=\{3,5,7,\ldots\}$, or fractional rates, e.g. $R=\log_2(257)$, are supported. Also, the centroid indexing is natural and simple. Extending the scope to correlated complex Gaussian sources, the circular design could be  transformed into an elliptical design,  $x_n=c_\textrm{re}\cos(\phi_n +\Delta \phi)+i\cdot c_\textrm{im}\sin(\phi_n +\Delta \phi), \, c_\textrm{re}\neq c_\textrm{im}$. In line with \cite{Larsson17b} on probabilistically-shaped modulation, we also recognize the opportunity for entropy encoded GQ designs. We finally note that the high-rate quantizer design given here (using high-rate quantization theory), and high-rate GB-GAM in \cite{Larsson17a, Larsson17b} (using inverse sampling method), have the same centroid and constellation point designs, respectively. This further substantiate the inverse sampling based approach in those works. The Lloyd-Max centroid design could be used for modulation. Such design improves the mutual information performance (not shown here) slightly over high-rate GB-GAM, but the PAPR improves with $\approx2.5$ dB, for $N=2^8$.

\subsection{Spiral Phyllotaxis, and Related Works}
Spiral phyllotaxis is the spiral arrangement of leaves (seeds, petals) on a plant. SP can be seen among, e.g., mosses (leafy shoot), ferns (leaf-branches), gymnosperm (e.g. cycad- and pine-cones), and angiosperm (e.g. sunflowers and dahlias). A historical source, on the study of SP, is  \cite{AdlerBaraJean97}. Vogel introduced the model $r_n=\sqrt{n}\mathrm{e}^{1i 2 \pi \varphi n}$ in \cite{Vogel79}. Many have noted SP in nature, and applied it in their work areas, e.g. antenna arrays in \cite{Boeringer00P}, orbital angular momenta of light \cite{LiewEtAl11}, and ultrasound imaging arrays in \cite{MartinezEtAl10}. In \cite{Larsson17a, Larsson17b}, the radial shape-flexibility of SP was recognized, and used, for geometrically-shaped golden angle modulation. Color pallet design for color \textit{Lab}-space, in shape of an irregular 3D-cone, was considered in \cite[Sec. III.B]{MojsilovicSolj01b}. Its cross-sections were uniformly sampled with SP, and invalid points, outside the irregular areas, were discarded. This gave irregularly clipped SPs, as shown in \cite[Fig.~5]{MojsilovicSolj01b}. In contrast to the present work, the independent color \textit{Lab}-space work does not; consider the problem of quantizing a continuous r.v. (with a pdf), use fully circular SP(s), recognize/use the shape-versatility of SP, nor adopt a non-uniform sampling distribution.

\section{Summary and Conclusions}
In this letter, we proposed non-uniform golden quantization for the complex Gaussian r.v., gave an analytical high-rate design with a rate-distortion expression, and a Lloyd-Max design. We found that the MSE improved compared to both polar-, and rectangular-, quantizer designs, and essentially coincided with the (best known) trained VQ designs. The proposed GQ also allowed for a structured design, a natural indexing order, and any integer number of centroids.

\bibliographystyle{IEEEtran}


\end{document}